\documentclass[aps, prd,  superscriptaddress, showpacs, longbibliography, floatfix, nofootinbib]{revtex4-2}

\usepackage{bm, amsmath, amssymb, relsize, amsfonts, mathrsfs, multirow, braket, siunitx, color, booktabs, arydshln, tabularx}
\usepackage{graphicx}
\graphicspath{{./figures/}}

\DeclareSIUnit \parsec {pc}

\usepackage{layouts}
\usepackage[dvipsnames]{xcolor}
\usepackage[unicode]{hyperref}
\hypersetup{
  colorlinks=true,
  citecolor=RoyalBlue,
  linkcolor=blue,
  urlcolor=blue
}

\usepackage{array}
\usepackage{cancel}
\usepackage{orcidlink}
\usepackage{graphicx}
\usepackage{dcolumn}
\usepackage{bm}
\usepackage{paralist}
\usepackage{epsfig}
\usepackage{placeins}
\usepackage{mathrsfs}
\usepackage{comment}
\usepackage{color, colortbl}
\usepackage[inline]{enumitem}
\usepackage{soul}
\usepackage{adjustbox}
\usepackage{amssymb}
\usepackage{caption}
\usepackage{subcaption}
\usepackage{float}
\usepackage{cancel}
\usepackage{verbatim}
\usepackage{amsmath}
\usepackage[toc,page]{appendix}
\usepackage{graphicx}
\usepackage{ragged2e}
\usepackage[T1]{fontenc}
\usepackage[utf8]{inputenc} 

\DeclareMathAlphabet{\mathpzc}{OT1}{pzc}{m}{it}
	
\definecolor{LightCyan}{rgb}{0.88,1,1}
\definecolor{lightgray}{gray}{0.9}

\def \IITGn     {Department of Physics, Indian Institute of Technology Gandhinagar, Gujarat 382355, India.\vspace*{4pt}}

\begin{document}

\title{Neutral Scalar Signatures at a Muon Collider in the $Z_3$ symmetric Three Higgs Doublet Model}

\author{\textsc{Baradhwaj Coleppa}\orcidlink{0000-0002-8761-3138}}
\email{baradhwaj@iitgn.ac.in }
\affiliation{\IITGn}

\author{\textsc{Akshat Khanna}\vspace*{7pt}\orcidlink{0000-0002-2322-5929}}
\email{khanna\textunderscore akshat@iitgn.ac.in}
\affiliation{\IITGn}

\begin{abstract}
Extending the scalar sector of the Standard Model is a well-motivated approach to exploring physics beyond the Standard Model. In this work, we investigate the phenomenology of the Three Higgs Doublet Model at a future muon collider. The scalar spectrum of the 3HDM comprises three CP-even Higgs bosons, two CP-odd Higgs bosons, and a pair of charged Higgs states. Focusing on Higgs pair production via muon-antimuon annihilation, we study the production and decay of neutral scalar states through the process $\mu^+\mu^- \to \phi_i \phi_j$, assuming a mass hierarchy in which the SM–like CP-even Higgs is the lightest state. We analyze several benchmark scenarios leading to $b\bar{b}b\bar{b}$ and $b\bar{b}t\bar{t}$ final states, and perform a cut-and-count analysis at a center-of-mass energy of $\sqrt{s}=3$ TeV. Our results demonstrate that a future muon collider provides a sensitive and promising environment to probe extended Higgs sectors, with neutral scalar states in the mass range of $200-400$ GeV being discoverable with $5\sigma$ significance for integrated luminosities of $\mathcal{O}(1$–$4 \ \mathrm{ab}^{-1})$.
\end{abstract}
\maketitle
\section{Introduction}
\label{sec:intro}

The Standard Model (SM) of particle physics~\cite{Salam:1968rm,Weinberg:1967tq,Glashow:1961tr}, based on the gauge symmetry group $SU(3)_c \times SU(2)_L \times U(1)_Y$, has achieved remarkable success in describing phenomena across a wide range of energy scales. The discovery of a Higgs boson with a mass near 125~GeV at the Large Hadron Collider (LHC) in 2012~\cite{Akeroyd_2021,CMS:2012qbp} completed the particle content predicted by the SM and marked a major milestone in our understanding of electroweak symmetry breaking. Despite this success, several fundamental questions remain unresolved. The SM does not provide an explanation for the origin of neutrino masses, the nature of dark matter, or the observed baryon asymmetry of the Universe. Moreover, many properties of the Higgs sector, particularly its self-interactions and couplings to fermions, remain only partially explored experimentally. These considerations provide strong motivation to investigate extensions of the scalar sector as a pathway to physics beyond the Standard Model (BSM).

A well-motivated and extensively studied class of such extensions involves enlarging the Higgs sector through additional $SU(2)_L$ scalar doublets. The Two Higgs Doublet Model (2HDM) and its various realizations~\cite{Branco:2011iw,Coleppa:2013dya,Chang:2012ve,Grinstein:2013npa,Drozd:2014yla,Bhattacharya:2023qfs,Muhlleitner:2016mzt,Keus:2017ioh,Bhattacharyya:2015nca,Cao:2009as} offer a rich phenomenology and have been widely explored in collider and flavor contexts~\cite{Ko:2013zsa,Crivellin:2013wna,PhysRevD.106.075003,PhysRevD.109.055016,PhysRevD.100.035031,Logan:2010ag}. Extending this idea further, the Three Higgs Doublet Model (3HDM) has attracted increasing attention in recent years~\cite{Cordero:2017owj,Ivanov:2012ry,Akeroyd:2021fpf,Boto:2025ovp,Ivanov:2014doa,Keus:2013hya}. By introducing two additional scalar doublets, the 3HDM predicts a significantly enriched scalar spectrum comprising multiple CP-even, CP-odd, and charged Higgs states. In contrast to the 2HDM, the 3HDM allows for a wider range of Yukawa structures and symmetry realizations, leading to qualitatively new phenomenological possibilities. In particular, scenarios with discrete symmetries can naturally enforce flavor structures that suppress tree-level flavor-changing neutral currents while simultaneously giving rise to distinctive scalar mixing patterns and decay signatures. Theoretical consistency and phenomenological implications of the 3HDM under various symmetry assumptions have been studied in detail in the literature~\cite{Batra:2025amk,Keus:2013hya,Coleppa:2025qst,PhysRevD.108.015020,PhysRevD.111.075009,PhysRevD.100.015008,Kalinowski:2021lvw,Das:2019yad,Chakraborti:2021bpy, Coleppa:2026ogl, Coleppa:2026fdj,Akeroyd:2016ssd,Romao:2025qag,Das:2025mqs,Kuncinas:2025uty,Kuncinas:2024zjq,Boto:2024jgj,Boto:2024tzp,Dey:2023exa,Das:2022gbm,Ivanov:2021pnr}.

The presence of additional scalar states provides a compelling target for experimental searches. Dedicated efforts at collider experiments, most notably at the LHC~\cite{ATLAS:2024itc,CMS:2024vps,CMS:2022suh,CMS-PAS-HIG-25-007,CMS-PAS-SUS-24-002,CMS:2024zfv,CMS:2024uru,CMS:2022fyt,ATLAS:2025qyn,ATLAS:2021ldb,ATLAS:2025rfm,CMS:2021yci}, have placed increasingly stringent constraints on the masses and couplings of non-SM Higgs bosons. The LHC currently operates at a center-of-mass energy of $\sqrt{s} = 13.6$~TeV, and its forthcoming high-luminosity phase (HL-LHC)~\cite{Azzi:2019yne,CidVidal:2018eel,ATLAS:2025eii} is expected to substantially improve sensitivity to extended Higgs sectors. However, searches for heavy scalars at hadron colliders are often limited by large QCD backgrounds and model-dependent production mechanisms, particularly in scenarios where couplings to gauge bosons are suppressed or where the dominant signals arise from cascade decays or associated production channels.

These challenges motivate the exploration of complementary experimental facilities. Future lepton colliders, such as the International Linear Collider (ILC)~\cite{Bambade:2019fyw,Dugan:2025utp,ILCInternationalDevelopmentTeam:2022izu,Abe:2025yur}, offer a clean experimental environment with reduced background levels, enabling precision measurements and direct pair production of new states up to roughly half the center-of-mass energy. At the same time, next-generation hadron machines such as the Future Circular Collider (FCC)~\cite{FCC:2025lpp,FCC:2025uan, Bernardi:2022hny,Benedikt:2022kan}, envisioned to operate at $\sqrt{s} \simeq 100$~TeV, promise an extended mass reach for new particles.

Within this broader landscape, the proposal of a high-energy multi-TeV muon collider~\cite{InternationalMuonCollider:2024jyv,InternationalMuonCollider:2025sys,Accettura:2023ked,Aime:2022flm} has emerged as a particularly powerful avenue for probing extended Higgs sectors. A muon collider combines the advantages of lepton collisions-clean initial states and well-defined kinematics-with access to very high center-of-mass energies. In contrast to hadron colliders, the full beam energy is available for the hard scattering process. Furthermore, owing to the larger muon mass, synchrotron radiation is strongly suppressed relative to electron–positron machines, allowing acceleration to multi-TeV energies with relatively small energy spread. These features make muon colliders especially well suited for the direct production of heavy scalar states and for detailed studies of their interactions. Recent studies have highlighted their exceptional potential for precision Higgs measurements, searches for heavy BSM particles, and probes of electroweak-scale dark matter candidates~\cite{Liu:2021jyc,Bao:2022onq,deLima:2025ctj,Han:2024gan,Frixione:2025guf,Lu:2023jlr,Capdevilla:2024bwt,Han:2021udl,Marin:2004hk,Das:2024ekt,Korshynska:2024suh,Cheung:2025uaz,Barik:2024kwv,Chakrabarty:2014pja,Ghosh:2023xbj,Ghosh:2025gdx}. The International Muon Collider Collaboration has proposed an initial stage operating at $\sqrt{s} = 3$~TeV with an integrated luminosity of $1000~\text{fb}^{-1}$, followed by a higher-energy stage at $\sqrt{s} = 10$~TeV with luminosities up to $10{,}000~\text{fb}^{-1}$.

In this work, we focus on a $Z_3$-symmetric realization of the Three Higgs Doublet Model with a Type-Z (democratic) Yukawa structure. This framework provides a natural implementation of Natural Flavor Conservation, with each fermion sector coupling to a different scalar doublet, thereby suppressing tree-level flavor-changing neutral currents while giving rise to a rich and predictive scalar sector. The resulting spectrum contains three CP-even Higgs bosons, two CP-odd Higgs bosons, and a pair of charged Higgs states, with multiple possible mass hierarchies and mixing patterns. Such a setup leads to a variety of novel collider signatures, particularly in channels involving associated production of CP-even and CP-odd scalars.

Here we perform a detector-level investigation of neutral scalar pair production in the $Z_3$-symmetric 3HDM at a future muon collider, providing a complementary probe of its extended Higgs sector. We investigate the collider phenomenology focusing on neutral scalar pair production via muon–antimuon annihilation $\mu^+\mu^- \to \phi_i \phi_j$ in the regular mass hierarchy where the SM-like CP-even Higgs boson is the lightest state. In this scenario, the heavier neutral scalars can be produced through associated $AH$ channels mediated by an off-shell $Z$ boson, providing a clean and direct probe of the extended Higgs sector. Representative benchmark points are selected that satisfy all relevant theoretical constraints, including vacuum stability, perturbative unitarity, and perturbativity, as well as current experimental bounds from Higgs signal strengths, direct searches, and flavor observables. We analyze final states arising from the subsequent decays of the heavy scalars, focusing on the $b\bar b b\bar b$ and $b\bar b t\bar t$ topologies. A cut-and-count analysis is performed at a center-of-mass energy of $\sqrt{s} = 3$~TeV. Our results show that a future muon collider provides a sensitive and promising environment for probing extended Higgs sectors and can achieve discovery-level significance for heavy neutral Higgs states over a substantial region of parameter space.

The structure of this paper is as follows. In Section~\ref{sec:model}, we describe the scalar content of the model and outline the key parameters and symmetries that define the $Z_3$-symmetric 3HDM. The Yukawa sector with Type-Z structure is then introduced, followed by a discussion of theoretical and experimental constraints in Section~\ref{sec:constraints}. In Section~\ref{sec:alignment}, we discuss the alignment limit under which one CP-even scalar reproduces the observed SM Higgs properties, and in Section~\ref{sec:couplings}, we present the relevant couplings. The collider phenomenology and signal significance analysis are presented in Section~\ref{sec:pheno}. Finally, we summarize our findings and provide an outlook in Section~\ref{sec:conclusions}.

\section{3HDM: Model}
\label{sec:model}
In this section, we provide a description of the specific 3HDM under consideration detailing the scalar and Yukawa sectors of the model. We also provide a table of the couplings relevant for the subsequent analysis. For a complete discussion of this model and its surviving parameter space after the imposition of all theoretical and experimental constraints, the reader is invited to consult Ref.~\cite{Batra:2025amk}.

\subsection{Scalar Sector}

In the framework of the 3HDM, the scalar sector of the SM is extended by introducing two additional $SU(2)_L$ doublets having exactly the same $U(1)_Y$ quantum numbers as that of the SM Higgs doublet. These can be conveniently represented in the usual fashion: 
\begin{equation}
		\Phi_k = \begin{pmatrix}
			\phi_k^+ \\ \frac{v_k+p_k+in_k}{\sqrt{2}}
		\end{pmatrix},
  \label{eq:phi_form}
\end{equation}
where $k=1,2,3$. We write the most general renormalizable scalar potential consistent with the $SU(2)_L\times U(1)_Y$ gauge symmetry and also is invariant under a discrete $Z_3$ symmetry:
\begin{equation}
     	\label{eq:scalarpot}
     	\begin{split}
     		V & = m_{11}^2(\Phi_1^\dagger\Phi_1) + m_{22}^2(\Phi_2^\dagger\Phi_2) + m_{33}^2(\Phi_3^\dagger\Phi_3)
     		\\ & + \lambda_1(\Phi_1^\dagger\Phi_1)^2 + \lambda_2(\Phi_2^\dagger\Phi_2)^2 + \lambda_3(\Phi_3^\dagger\Phi_3)^2 \\ & + \lambda_{4}(\Phi_1^\dagger\Phi_1)(\Phi_2^\dagger\Phi_2) + \lambda_{5}(\Phi_1^\dagger\Phi_1)(\Phi_3^\dagger\Phi_3) + \lambda_{6}(\Phi_2^\dagger\Phi_2)(\Phi_3^\dagger\Phi_3) \\ & + \lambda_{7}(\Phi_1^\dagger\Phi_2)(\Phi_2^\dagger\Phi_1) + \lambda_{8}(\Phi_1^\dagger\Phi_3)(\Phi_3^\dagger\Phi_1) +  \lambda_{9}(\Phi_2^\dagger\Phi_3)(\Phi_3^\dagger\Phi_2) \\ & + [\lambda_{10}(\Phi_1^\dagger\Phi_2)(\Phi_1^\dagger\Phi_3) + \lambda_{11}(\Phi_1^\dagger\Phi_2)(\Phi_3^\dagger\Phi_2) + \lambda_{12}(\Phi_1^\dagger\Phi_3)(\Phi_2^\dagger\Phi_3) + h.c.]. \\ &
     	\end{split} 
\end{equation}
Under the additional $Z_3$ symmetry, the Higgs fields transform as
\begin{equation}
     	\label{eq:phitrans}
     	\Phi_1 \rightarrow \omega \Phi_1, \; \; \Phi_2 \rightarrow \omega^2 \Phi_2, \;\textrm{and} \;\Phi_3 \rightarrow \Phi_3,
\end{equation}
where $\omega = e^{2\pi i/3}$ are the cube roots of unity. In general the parameters $\lambda_{1,2,...,9}$ of the potential are real (owing to the hermiticity of the Lagrangian), while $\lambda_{10},\lambda_{11}$ and $\lambda_{12}$ can be complex. To prevent CP-even and CP-odd scalar mixing, we restrict our analysis to a CP-conserving scenario by setting all complex parameters in the potential to zero. We begin with the  relevant mass terms for the CP-even Higgs bosons that are readily deduced from the scalar potential in Eqn.~\ref{eq:scalarpot} and can be written as
\begin{equation*}
		V_p^{mass} \supset \begin{pmatrix}
			p_1 & p_2 & p_3
		\end{pmatrix} \frac{\mathcal{M}^2_S}{2} \begin{pmatrix}
			p_1 \\ p_2 \\ p_3 
		\end{pmatrix},
\end{equation*}
where the elements of the mass matrix $\mathcal{M}_S^2$ are given by,
\begin{equation}
    \begin{split}
        (\mathcal{M}^2_S)_{11} & = 2\lambda_{1}v_1^2 - \frac{v_2v_3}{2v_1}[Re(\lambda_{11})v_2+Re(\lambda_{12})v_3]\\     (\mathcal{M}^2_S)_{12} & = (\lambda_{4}+\lambda_{7})v_1v_2 + Re(\lambda_{10})v_1v_3 + Re(\lambda_{11})v_2v_3 + \frac{v_3^2}{2}Re(\lambda_{12})\\
        (\mathcal{M}^2_S)_{13} & = (\lambda_{5}+\lambda_{8})v_1v_3 + Re(\lambda_{10})v_1v_2 + Re(\lambda_{12})v_2v_3 + \frac{v_2^2}{2}Re(\lambda_{11})\\
        (\mathcal{M}^2_S)_{22} & =  2\lambda_{2}v_2^2 - \frac{v_1v_3}{2v_2}[Re(\lambda_{10})v_1+Re(\lambda_{12})v_3] \\
        (\mathcal{M}^2_S)_{23} & = (\lambda_{6}+\lambda_{9})v_2v_3 + Re(\lambda_{11})v_1v_2 + Re(\lambda_{12})v_1v_3 + \frac{v_1^2}{2}Re(\lambda_{10}) \\
        (\mathcal{M}^2_S)_{33} & = 2\lambda_{3}v_3^2 - \frac{v_1v_2}{2v_3}[Re(\lambda_{10})v_1+Re(\lambda_{11})v_2]
    \end{split}
    \label{eq:scalarmassterms}
\end{equation}	
This real symmetric mass matrix can be diagonalized by an orthogonal transformation by a matrix $O_\alpha$ defined as
\begin{equation}
    \label{eq:matalphtrans}
    O_\alpha = \begin{pmatrix}
        c_{\alpha 1} c_{\alpha 2} & c_{\alpha 2} s_{\alpha 1} & s_{\alpha 2} \\ -c_{\alpha 3} s_{\alpha 1} - s_{\alpha 3} s_{\alpha 2} c_{\alpha 1} & c_{\alpha 3} c_{\alpha 1} - s_{\alpha 3} s_{\alpha 2} s_{\alpha 1} & s_{\alpha 3} c_{\alpha 2} \\ s_{\alpha 3} s_{\alpha 1} - c_{\alpha 3} s_{\alpha 2} c_{\alpha 1} & -s_{\alpha 3} c_{\alpha 1} - c_{\alpha 3} s_{\alpha 2} s_{\alpha 1} & c_{\alpha 3} c_{\alpha 2} 
    \end{pmatrix}.
\end{equation}
The diagonalization $O_\alpha\mathcal{M}^2_S O_\alpha^T$ yields the masses of the three CP-even Higgs bosons in the model which we designate as $m_{H 1}^2$, $m_{H 2}^2$, and $m_{H 3}^2$.
The scalar masses $m_{Hi}^2$  are functions of the couplings $\lambda_i$s in the potential. In this work, we consider the scalar masses and mixing angles as independent parameters, and compute the $\lambda_i's$ in terms of them. It is clear from Eqn.~\ref{eq:matalphtrans} that the mass eigenstates can be written down in terms of the gauge eigenstates in a straightforward manner: 
\begin{equation}
    \begin{split}
        H_1 & = c_{\alpha_2} c_{\alpha_1} p_1 + c_{\alpha_2} s_{\alpha_1} p_2 + s_{\alpha_2}p_3, \\
        H_2 & = -(c_{\alpha_3} s_{\alpha_1} + s_{\alpha_3} s_{\alpha_2} c_{\alpha_1})p_1 + (c_{\alpha_3} c_{\alpha_1} - s_{\alpha_3} s_{\alpha_2} s_{\alpha_1})p_2+(s_{\alpha_3} c_{\alpha_2})p_3,\,\textrm{and} \\
        H_3 & =  (s_{\alpha_3} s_{\alpha_1} - c_{\alpha_3} s_{\alpha_2} c_{\alpha_1})p_1 -  (s_{\alpha_3} c_{\alpha_1} + c_{\alpha_3} s_{\alpha_2} s_{\alpha_1})p_2 + (c_{\alpha_3} c_{\alpha_2}) p_3.
    \end{split}
\end{equation}
The mass terms for the charged Higgses can similarly be extracted from the scalar potential given in Eqn.~\ref{eq:scalarpot} and can be symbolically written as
\begin{equation*}
     	V_C^{mass} \supset \begin{pmatrix}
     		\phi_1^- & \phi_2^- & \phi_3^-
     	\end{pmatrix} \mathcal{M}^2_{\phi^{\pm}} \begin{pmatrix}
     		\phi_1^+ \\ \phi_2^+ \\ \phi_3^+ 
     	\end{pmatrix},
\end{equation*}
where, $\mathcal{M}^2_{\phi^{\pm}}$ is the $3 \times 3$ charged Higgs mass matrix whose elements are given below:
\begin{equation*}
    \begin{split}
        (\mathcal{M}^2_{\phi^{\pm}})_{11} & = -\frac{v_2^2}{2}\lambda_{7} - \frac{v_3^2}{2}\lambda_{8} - Re(\lambda_{10})v_2v_3 - \frac{v_2v_3}{2v_1}(Re(\lambda_{11})v_2+Re(\lambda_{12})v_3), \\  (\mathcal{M}^2_{\phi^{\pm}})_{12} & = \frac{v_1v_2}{2}\lambda_{7} + \frac{v_1v_3}{2}\lambda_{10} + \frac{v_2v_3}{2}\lambda_{11},\\
        (\mathcal{M}^2_{\phi^{\pm}})_{13} & = \frac{v_1v_2}{2}\lambda_{10} + \frac{v_1v_3}{2}\lambda_{8} + \frac{v_2v_3}{2}\lambda_{12},\\
        (\mathcal{M}^2_{\phi^{\pm}})_{22} & = -\frac{v_1^2}{2}\lambda_{7} - \frac{v_3^2}{2}\lambda_{9} - Re(\lambda_{11})v_1v_3 - \frac{v_1v_3}{2v_2}(Re(\lambda_{10})v_1+Re(\lambda_{12})v_3), \\
        (\mathcal{M}^2_{\phi^{\pm}})_{23} & = \frac{v_1v_2}{2}\lambda_{11}^* + \frac{v_1v_3}{2}\lambda_{12} + \frac{v_2v_3}{2}\lambda_{9},\,\,\textrm{and} \\
        (\mathcal{M}^2_{\phi^{\pm}})_{33} & = -\frac{v_1^2}{2}\lambda_{8} - \frac{v_2^2}{2}\lambda_{9} - Re(\lambda_{12})v_1v_2 - \frac{v_1v_2}{2v_3}(Re(\lambda_{10})v_1+Re(\lambda_{11})v_2).
    \end{split}
\end{equation*}
To diagonalize this, we first employ a similarity transformation using the matrix $O_\beta$: 
\begin{equation*}
    (B_C)^2 = O_\beta.\mathcal{M}^2_{\phi^{\pm}}.O_\beta^T,
\end{equation*}
where
\begin{eqnarray}
    \label{eq:betatrans}
    O_\beta &=
     & \begin{pmatrix}
        c_{\beta 2}c_{\beta 1} & c_{\beta 2}s_{\beta 1} & s_{\beta 2} \\ -s_{\beta 1} & c_{\beta 1} & 0 \\ -c_{\beta 1}s_{\beta 2} & -s_{\beta 1}s_{\beta 2} & c_{\beta 2}
    \end{pmatrix}
\end{eqnarray}
Here, $\tan\beta_1=v_2/v_1$ and $\tan\beta_2=v_3/\sqrt{v_1^2+v_2^2}$. This similarity transformation brings the matrix to a block diagonal form and this can be fully diagonalized by a subsequent transformation with a matrix $O_\gamma$ given by
\begin{equation}
    \label{gam2trans}
    O_{\gamma 2} =  \begin{pmatrix} 1 & 0 & 0 \\ 0 & c_{\gamma 2} & -s_{\gamma 2} \\ 0 & s_{\gamma 2} & c_{\gamma 2}
    \end{pmatrix}
\end{equation}
and the final form explicitly has a zero eigenvalue corresponding to the Goldstone boson eaten by the $W^\pm$:
\begin{equation*}
     	O_{\gamma 2}.(B_C)^2.O_{\gamma 2}^{T} = \begin{pmatrix}
     		0 & 0 & 0 \\ 0 & m_{H^{\pm}_2}^2 & 0 \\ 0 & 0 & m_{H^{\pm}_3}^2
     	\end{pmatrix}.
\end{equation*}
As in the case of CP-even scalars, analogous relations can be derived connecting the parameters $\lambda_7,\lambda_8,\lambda_9$ to the charged Higgs masses, mixing angles and the vevs. The gauge eigenstates can hence be represented in terms of mass eigenstates as:
\begin{equation}
    \label{chargetransrelat}
    \begin{split}
        G^\pm & = c_{\beta 1}c_{\beta 2} \phi_1^\pm + c_{\beta 2}s_{\beta 1} \phi_2^\pm + s_{\beta 2} \phi_3^\pm, \\
        H_2^\pm & = (-c_{\gamma 2 }s_{\beta 1} + c_{\beta 1}s_{\beta 2}s_{\gamma 2})\phi_1^\pm + (c_{\beta 1}c_{\gamma 2} + s_{\beta 1}s_{\beta 2}s_{\gamma 2})\phi_2^\pm + (-c_{\beta 2}s_{\gamma 2}) \phi_3^\pm,\,\textrm{and} \\
        H_3^\pm & = (-c_{\beta 1}c_{\gamma 2}s_{\beta 2}-s_{\beta 1}s_{\gamma 2})\phi_1^\pm + (-c_{\gamma 2}s_{\beta 1}s_{\beta 2}+c_{\beta 1}s_{\gamma 2})\phi_2^\pm + (c_{\beta 2}c_{\gamma 2}) \phi_3^\pm.
    \end{split}
\end{equation}
Finally, writing the mass terms for the CP-Odd Higgs in a similar fashion
\begin{equation*}
        V_n^{mass} \supset \begin{pmatrix}
            n_1 & n_2 & n_3
        \end{pmatrix} \frac{\mathcal{M}^2_n}{2} \begin{pmatrix}
            n_1 \\ n_2 \\ n_3 
        \end{pmatrix}, 
\end{equation*}
we diagonalize the pseudoscalar mass matrix exactly like in the previous case, \textit{i.e}., perform two consecutive rotations, first by $O_\beta$ and then by $O_{\gamma1}$. Once again, we trade the Lagrangian parameters with masses, mixing angles, and vevs. The mass eigenstates can hence be represented in terms of gauge eigenstates as
\begin{equation}
    \begin{split}
        G_0 & = (c_{\beta 1}c_{\beta 2}) n_1 + (c_{\beta 2}s_{\beta 1}) n_2 + (s_{\beta 2}) n_3, \\
        A_1 & = (-c_{\gamma 1 }s_{\beta 1} + c_{\beta 1}s_{\beta 2}s_{\gamma 1})n_1 + (c_{\beta 1}c_{\gamma 1} + s_{\beta 1}s_{\beta 2}s_{\gamma 1})n_2 + (-c_{\beta 2}s_{\gamma 1})n_3,\,\textrm{and} \\
        A_2 & = (-c_{\beta 1}c_{\gamma 1}s_{\beta 2}-s_{\beta 1}s_{\gamma 1})n_1 + (-c_{\gamma 1}s_{\beta 1}s_{\beta 2}+c_{\beta 1}s_{\gamma 1})n_2 + (c_{\beta 2}c_{\gamma 1}) n_3.
    \end{split}
\end{equation}
\subsection{Yukawa Sector}
Tree-level Flavor-Changing Neutral Currents (FCNCs) pose a significant challenge to multi-Higgs doublet models due to stringent experimental limits. To avoid such currents, one of two avenues is commonly pursued: all the fermion Yukawas arise from only one of the Higgs doublets, or each doublet couples to distinct fermions. In this work, we implement Natural Flavor Conservation (NFC), following the second route \cite{Batra:2025amk} which mandates that each class of fermions couples to a single Higgs doublet. This prevents the appearance of FCNCs at tree level by construction. To realize this condition, we adopt the Type-Z Yukawa structure, also referred to as the democratic scenario. In this setup, the up-type quarks, down-type quarks, and the charged leptons couple to one Higgs doublet each ensuring that mass generation occurs independently for each sector. The resulting Yukawa Lagrangian for the model is given by
\begin{equation}
		\label{eq:yukeq}
		\mathcal{L}_{Yukawa} = -[\bar{L}_L \Phi_1 \mathcal{G}_l l_R+\bar{Q}_L \Phi_2 \mathcal{G}_d d_R + \bar{Q}_L \tilde{\Phi}_3 \mathcal{G}_u u_R  + h.c],
\end{equation}
where the $\mathcal{G}_f$ are the Yukawa matrices. In terms of the fermion mass matrices they can be written as
\begin{equation*}
		\mathcal{G}_f = \frac{\sqrt{2} \mathcal{M}_f}{v_i}.
\end{equation*}
We work with a $Z_3$ symmetric potential as given in Eqn.~\ref{eq:phitrans} - for the Yukawa Lagrangian to remain invariant under the same, the right handed fermion fields transform as 
\begin{equation}
		\label{eq:fermtrans}
		d_R  \rightarrow \omega d_R , \; \; \; \; l_R \rightarrow \omega^2 l_R, \; \; \; \; u_R \rightarrow  u_R.
\end{equation}
It is instructive to contrast this setup with the familiar Yukawa structures of the 2HDM, in particular the Type-II scenario. In the Type-II 2HDM, up-type quarks couple to one scalar doublet, while down-type quarks and charged leptons couple to the other, thereby enforcing Natural Flavor Conservation but correlating the mass generation of two fermion sectors. In the Type-Z realization of the 3HDM, as explained above, this structure is generalized by assigning a separate Higgs doublet to each fermion class: up-type quarks, down-type quarks, and charged leptons couple to different scalar doublets. This leads to a more flexible and less correlated Yukawa sector, with two independent ratios of vacuum expectation values and a richer pattern of scalar–fermion couplings. As a result, the resulting decay hierarchies and collider signatures can differ qualitatively from those of the 2HDM, providing additional motivation to study this framework in dedicated search strategies.

\section{Constraints}
\label{sec:constraints}
The issue of constraining the democratic 3HDM imposing all theoretical and experimental results and extracting the surviving parameter space of the model was dealt with in detail in Ref.~\cite{Batra:2025amk}. Here, we give a brief overview of the various constraints. The theoretical consistency of the model is ensured by demanding vacuum stability, unitarity, and perturbativity. The vacuum stability requirement ensures that the scalar potential is bounded from below in all directions of the field space, thereby preventing the emergence of unstable vacua while the perturbativity constraints (which are upper bounds on the various couplings, specifically, $|\lambda_i| \leq 4\pi$ for all $i$) are imposed to ensure that the theory remains perturbatively calculable. Following the results of Ref.~\cite{Bento:2022vsb}, we apply the unitarity conditions derived for the $Z_3$-symmetric 3HDM, requiring that all 21 eigenvalues $\Lambda_i$ of the relevant scalar scattering matrices satisfy $|\Lambda_i| \leq 8\pi$. 

In addition to theoretical requirements, our model is subjected to several experimental constraints to ensure its consistency with current data. The electroweak precision tests play a crucial role in probing the effects of new physics on gauge boson propagators through the oblique parameters $S$, $T$, and $U$. These parameters impose stringent restrictions on extensions of the SM, with their experimentally measured values given by \cite{Workman:2022ynf}
\begin{align*}
        S & = -0.02 \pm 0.10, \\
        T & = 0.03 \pm 0.12,\, \textrm{and} \\
        U & = 0.01 \pm 0.11. 
\end{align*}
Furthermore, the BSM Higgs boson exclusion limits are evaluated using direct search results from LEP, Tevatron, and the LHC. These constraints are applied at the $95\%$ Confidence Level (C.L.) using the HiggsBounds-6 framework, interfaced through the HiggsTools package \cite{Bahl_2023}. To ensure the SM-like Higgs boson discovery is accurately reproduced, the compatibility of the 125 GeV Higgs state in the present model is assessed with experimental observations through a goodness-of-fit test. Finally, flavour physics constraints are incorporated by imposing the most stringent bounds on the branching ratio $\mathcal{BR}(B \rightarrow X_s \gamma)$, computed using next-to-leading order (NLO) QCD corrections as discussed in \cite{Akeroyd_2021,Boto_2021}. The following restriction has been imposed which represents the $3 \sigma$ experimental limit: 
\begin{equation*}
    2.87 \times 10^{-4} < \mathcal{BR}(B \rightarrow X_s \gamma) < 3.77 \times 10^{-4}.
\end{equation*}

\section{Alignment Limit in the 3HDM}
\label{sec:alignment}
The observation of a 125 GeV Higgs boson with properties closely matching those predicted by the SM implies that one of the three CP-even Higgs bosons in the 3HDM must play the role of the SM-like Higgs. Requiring this defines the so-called alignment limit, a condition under which one of the CP-Even Higgs states acquires SM like couplings. A distinctive feature of the 3HDM is the absence of a predetermined mass hierarchy among the CP-even Higgs states; their ordering depends sensitively on the numerical values of the quartic couplings ($\lambda_i$) and the vacuum expectation values (VEVs). This flexibility allows for multiple possible hierarchies between the three CP-even mass eigenstates, wherein any of the three CP-even states could, in principle, serve as the SM-like Higgs boson. In this work, we consider the following scenario:

\textbf{Regular Hierarchy:} The lightest CP-even scalar is identified as the SM-like Higgs. In this scenario, $H_1$ is identified \footnote{Note that the absence of a pre-determined mass hierarchy means that we can take any one of $H_1$, $H_2$, or $H_3$ to be the 125 GeV Higgs - fixing this as $H_1$ like we have done here constitutes a specific choice.} as the SM–like Higgs, while the remaining two CP-even Higgs states are taken to be heavy. The alignment limit condition for the regular hierarchy then implies that $g_{H_1ZZ}$ be SM-like which translates into 
 \begin{equation}   	       c_{\beta_2}c_{\alpha_2}\cos(\alpha_1-\beta_1) + s_{\beta_2}s_{\alpha_2} = 1.
      \label{eq:al1}
 \end{equation}
 Letting $k=\cos{(\alpha_1-\beta_1)}$, this condition is satisfied if $k=1\implies \alpha_1 = \beta_1 + 2n\pi$ and  $\alpha_2 = \beta_2 + 2n\pi$.
For $k=-1$, the scenario is equivalent to the present case, modulo an inversion. In contrast, $k\neq 1$ corresponds to a distinct configuration in which the full vacuum expectation value (vev) is localized within the third doublet. Accordingly, we restrict our subsequent analysis to the case $k=1$ \cite{Batra:2025amk}. \\

\begin{figure}[h!]
        \includegraphics[scale=0.7]{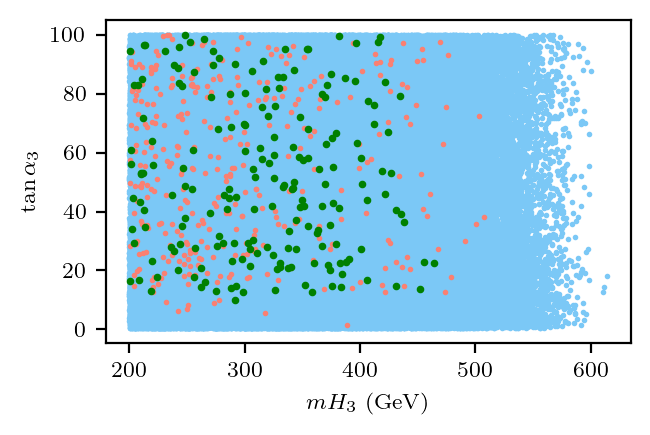}
        \includegraphics[scale=0.7]{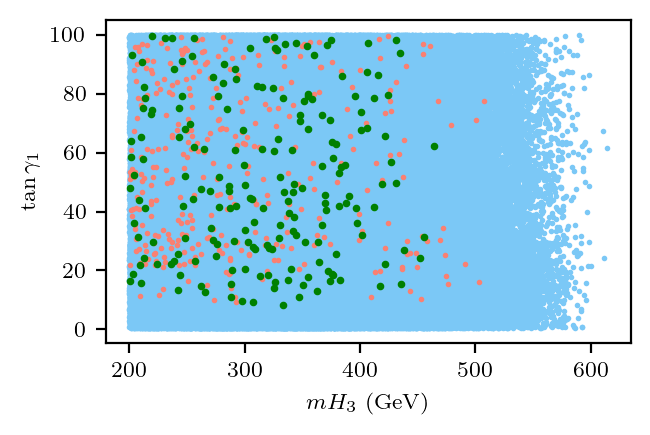}
        
        \caption{\justifying Allowed parameter space for the mixing angles $\gamma_1$ and $\alpha_3$. The color coding of the plots is the same as that in Figure \ref{fig:mH2-al1-masses}. }
         \label{fig:mH2-al1-angles}
\end{figure}

\begin{figure}[h!]
        \includegraphics[scale=0.7]{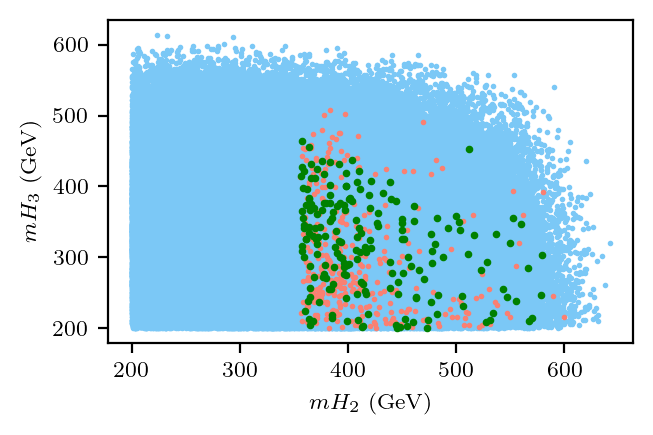}
        \includegraphics[scale=0.7]{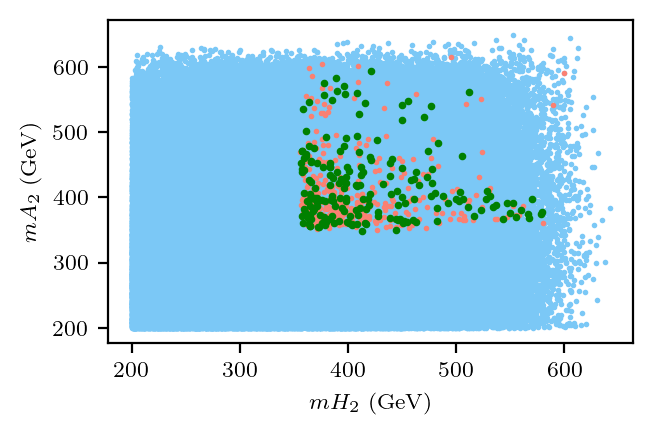}
        \includegraphics[scale=0.7]{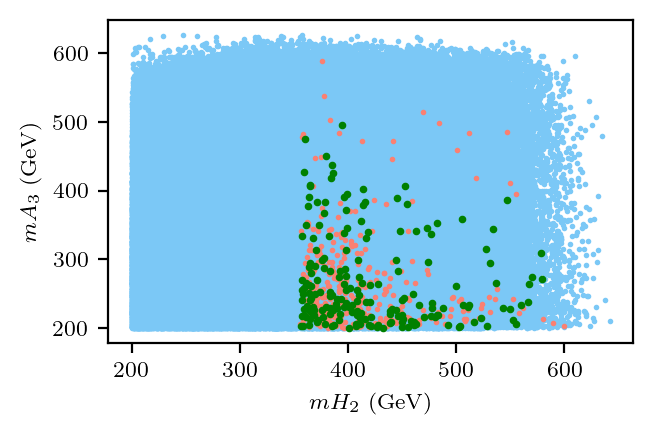}
        \includegraphics[scale=0.7]{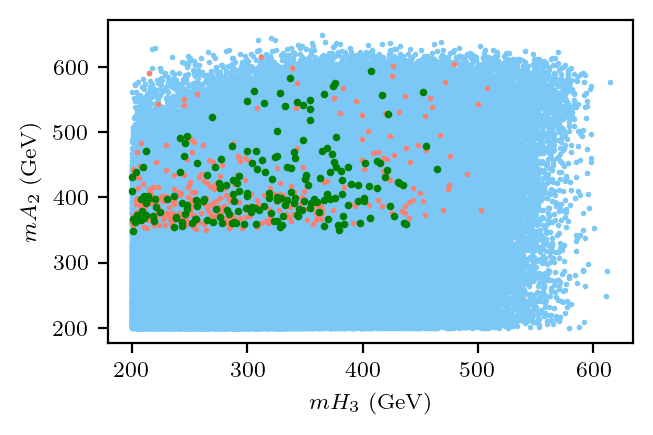}
        \includegraphics[scale=0.7]{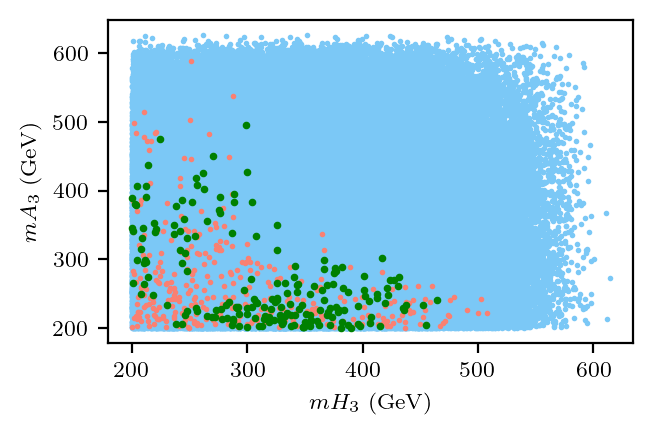}
        \includegraphics[scale=0.7]{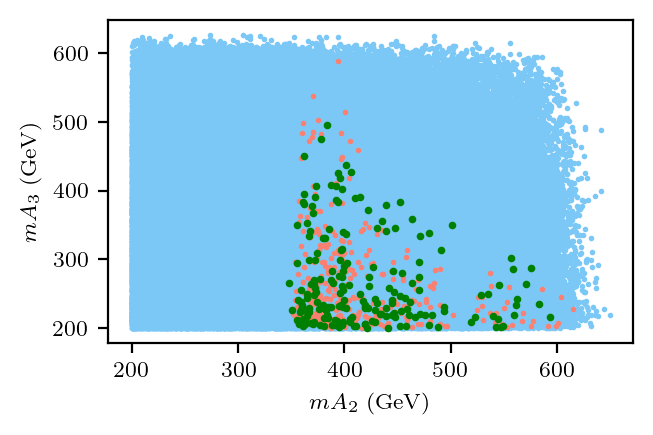}
        
        \caption{\justifying Allowed parameter space for the regular mass hierarchy scenario, for the neutral scalars, where the lightest CP-even scalar is identified with the 125 GeV SM-like Higgs boson. The figure presents the allowed regions for the masses of all heavy neutral Higgs states and their correlations with the relevant scalar mixing angles. The blue shaded regions satisfy the theoretical constraints from vacuum stability, perturbativity, and unitarity. The red regions additionally comply with direct search limits, goodness-of-fit requirements, and the flavor constraint from $b \rightarrow s \gamma$. Furthermore, the parameter points also consistent with electroweak precision observables are shown in green.}
         \label{fig:mH2-al1-masses}
\end{figure}

Figures \ref{fig:mH2-al1-angles} and \ref{fig:mH2-al1-masses} (reproduced from \cite{Batra:2025amk}) illustrate the viable parameter space for the masses of the additional Higgs bosons and their correlations with the mixing angles $\gamma_1$ and $\alpha_3$, which govern CP-odd and CP-even scalar mixing, respectively. The blue shaded region satisfies the theoretical requirements of vacuum stability, perturbativity, and unitarity. Imposing experimental bounds further restricts the parameter space to the red region, while the subset consistent with electroweak precision observables is highlighted in green. Consequently, the green region represents the parameter space simultaneously compatible with all theoretical, experimental, and electroweak precision constraints. From the figure, one observes that the mass of $H_2$ is strongly constrained to lie in the range $350~\text{GeV} < m_{H_2} < 580~\text{GeV}$, whereas $H_3$ is allowed within $200~\text{GeV} < m_{H_3} < 460~\text{GeV}$. The permitted mass intervals for $A_2$ and $A_3$ are found to be similar in size. In contrast, the scalar mixing angles $\gamma_1$ and $\alpha_3$ are only weakly constrained by the imposed bounds. Nevertheless, the viable parameter space shows a clear preference for large mixing angles, with $\tan\alpha_3 \gtrsim 10$ and $\tan\gamma_1 \gtrsim 8.2$, corresponding to $\alpha_3, \gamma_1 \gtrsim 84^\circ$.
\section{Couplings}
\label{sec:couplings}
In this section, we evaluate all relevant interaction vertices. The $AHZ$-type couplings are collected in Table~\ref{tab:couplings}, with $k_2 = \sin(\alpha_1 - \beta_1)$. Each coupling is expressed up to a common overall factor of $(e/2)\csc\theta_W \sec\theta_W$. For completeness, we also provide the corresponding simplified expressions obtained after imposing the alignment limit. In this limit, the coupling between the CP-odd Higgs boson, the SM-like Higgs, and the $Z$ boson vanishes identically. In contrast, the couplings involving the CP-odd state and the remaining two CP-even Higgs bosons display a complementary pattern: an enhancement of one coupling is accompanied by a suppression of the other. This anticorrelation originates purely from their dependence on the mixing angles $\gamma_1$ and $\alpha_3$.
\begin{table}[h!]
    \centering
    \begin{tabular}{l l l l}
    \toprule[1pt]
        Coupling & \ \ \ \ \ Coupling & \ \ \ \ \ Regular  \\
        \midrule[1pt]
        $A_2H_1Z$ & \ \ \ \ \  $c_{\beta_2}s_{\alpha_2}s_{\gamma_1}-c_{\alpha_2}(c_{\gamma_1}k_2+s_{\beta_2}s_{\gamma_1}k_1)$ & \ \ \ \ \ $0$  \\
        $A_2H_2Z$ & \ \ \ \ \  $c_{\alpha_3}(-c_{\gamma_1}k_1+s_{\beta_2}s_{\gamma_1}k_2) + s_{\alpha_3}(c_{\gamma_1}s_{\alpha_2}k_2+s_{\gamma_1}(c_{\alpha_2}c_{\beta_2}+s_{\alpha_2}s_{\beta_2}k_1))$ & \ \ \ \ \ $- c_{(\gamma_1+\alpha_3)}$  \\
        $A_2H_3Z$ & \ \ \ \ \  $-s_{\alpha_3}s_{\beta_2}s_{\gamma_1}k_2 + c_{\alpha_3}(s_{\alpha_2}c_{\gamma_1}k_2+c_{\alpha_2}c_{\beta_2}s_{\gamma_1}) + k_1(s_{\alpha_3}c_{\gamma_1}+c_{\alpha_3}s_{\alpha_2}s_{\beta_2}s_{\gamma_1})$ & \ \ \ \ \ $s_{(\gamma_1+\alpha_3)}$  \\
        $A_3H_1Z$ & \ \ \ \ \  $-(c_{\beta_2}s_{\alpha_2}c_{\gamma_1}+c_{\alpha_2}(-s_{\beta_2}c_{\gamma_1}k_1+s_{\gamma_1}k_2))$ & \ \ \ \ \ $0$  \\
        $A_3H_2Z$ & \ \ \ \ \  $-(c_{\alpha_2}c_{\beta_2}c_{\gamma_1}s_{\alpha_3}+k_1(s_{\alpha_2}s_{\beta_2}s_{\alpha_3}c_{\gamma_1}+s_{\gamma_1}c_{\alpha_3}) + k_2(s_{\beta_2}c_{\alpha_3}c_{\gamma_1}-s_{\alpha_2}s_{\alpha_3}s_{\gamma_1}))$ & \ \ \ \ \ $-s_{(\gamma_1+\alpha_3)}$  \\
        $A_3H_3Z$ & \ \ \ \ \  $-c_{\alpha_2}c_{\beta_2}c_{\alpha_3}c_{\gamma_1} + s_{\alpha_3}(s_{\beta_2}c_{\gamma_1}k_2+s_{\gamma_1}k_1)+c_{\alpha_3}s_{\alpha_2}(-s_{\beta_2}c_{\gamma_1}k_1+s_{\gamma_1}k_2)$ & \ \ \ \ \ $- c_{(\gamma_1+\alpha_3)}$  \\
        ${\bar{b} b} {H_1}$ & \ \ \ \ \  $\frac{s_{\alpha_1}c_{\alpha_2}}{s_{\beta_1}c_{\beta_2}}$ &  \ \ \ \ \ 1 \\
        ${\bar{b} b} {H_2}$ & \ \ \ \ \  $-\frac{(c_{\alpha_1}c_{\alpha_3}-s_{\alpha_1}s_{\alpha_2}s_{\alpha_3})}{s_{\beta_1}c_{\beta_2}}$ & \ \ \ \ \ $-\frac{(c_{\beta_1}c_{\alpha_3}-s_{\beta_1}s_{\beta_2}s_{\alpha_3})}{s_{\beta_1}c_{\beta_2}}$ \\
        ${\bar{b} b} {H_3}$ & \ \ \ \ \  $\frac{(c_{\alpha_1}s_{\alpha_3}+s_{\alpha_1}s_{\alpha_2}c_{\alpha_3})}{s_{\beta_1}c_{\beta_2}}$ & \ \ \ \ \ $\frac{(c_{\beta_1}s_{\alpha_3}+s_{\beta_1}s_{\beta_2}c_{\alpha_3})}{s_{\beta_1}c_{\beta_2}}$ \\
        ${\bar{t} t} {H_1}$ & \ \ \ \ \  $\frac{s_{\alpha_2}}{s_{\beta_2}}$ & \ \ \ \ \ 1 \\
        ${\bar{t} t} {H_2}$ & \ \ \ \ \  $\frac{c_{\alpha_2}s_{\alpha_3}}{s_{\beta_2}}$ & \ \ \ \ \ $\frac{c_{\beta_2}s_{\alpha_3}}{s_{\beta_2}}$ \\
        ${\bar{t} t} {H_3}$ & \ \ \ \ \  $\frac{c_{\alpha_2}c_{\alpha_3}}{s_{\beta_2}}$ & \ \ \ \ \ $\frac{c_{\beta_2}c_{\alpha_3}}{s_{\beta_2}}$ \\
        $\bar{b} b A_2$ & \ \ \ \ \ $\frac{c_{\beta_1}c_{\gamma_1}+s_{\beta_1}s_{\beta_2}s_{\gamma_1}}{s_{\beta_1}c_{\beta_2}}$ & \ \ \ \ \ $\frac{c_{\beta_1}c_{\gamma_1}+s_{\beta_1}s_{\beta_2}s_{\gamma_1}}{s_{\beta_1}c_{\beta_2}}$ \\
        $\bar{b} b A_3$ & \ \ \ \ \ $\frac{-c_{\beta_1}s_{\gamma_1}+s_{\beta_1}s_{\beta_2}c_{\gamma_1}}{s_{\beta_1}c_{\beta_2}}$ & \ \ \ \ \ $\frac{-c_{\beta_1}s_{\gamma_1}+s_{\beta_1}s_{\beta_2}c_{\gamma_1}}{s_{\beta_1}c_{\beta_2}}$ \\
        $\bar{t} t A_2$ & \ \ \ \ \ $\frac{c_{\beta_2}s_{\gamma_1}}{s_{\beta_2}}$ & \ \ \ \ \ $\frac{c_{\beta_2}s_{\gamma_1}}{s_{\beta_2}}$ \\
        $\bar{t} t A_3$ & \ \ \ \ \ $\frac{c_{\beta_2}c_{\gamma_1}}{s_{\beta_2}}$ & \ \ \ \ \ $\frac{c_{\beta_2}c_{\gamma_1}}{s_{\beta_2}}$ \\
    \bottomrule[1pt]
    \bottomrule[1pt]
    \end{tabular}
    \caption{ \justifying The table lists all relevant $AHZ$ couplings along with the couplings of the neutral scalar states to quarks, normalized to the corresponding SM Higgs couplings. The third column presents these couplings in the regular hierarchy scenario after imposing the alignment limit.}
    \label{tab:couplings}
\end{table}
\section{Collider Phenomenology}
\label{sec:pheno}
\begin{figure}[h!]
    \centering
    \includegraphics[scale=0.5]{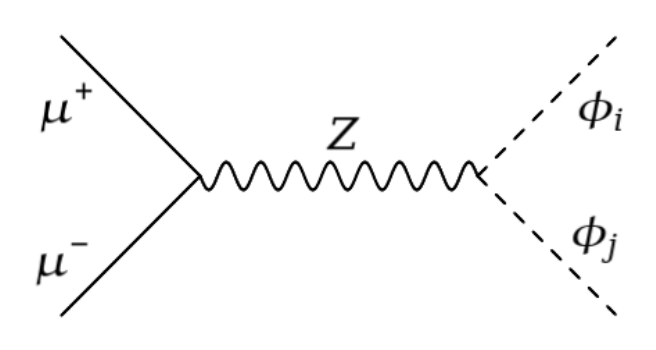}    
    \caption{Feynman diagram illustrating the scalar boson pair production through $\mu^+\mu^-$ annihilation. }
    \label{fig:feynman}
\end{figure}

With all the pieces in place, we are now ready to explore the collider search prospects of the neutral scalar states at a muon collider. The $\mu^+\mu^-$ annihilation channel offers an efficient avenue for Higgs pair production, as it enables optimal use of the available center of mass energy for producing heavy states. The corresponding Feynman diagram is shown in Fig.~\ref{fig:feynman}, where $\phi_i$ represents any neutral scalar mass eigenstate, either CP-even or CP-odd.
In the 3HDM, the physical scalar spectrum comprises three CP-even Higgs bosons, $H_1$, $H_2$, and $H_3$, along with two CP-odd Higgs bosons, $A_2$ and $A_3$. This rich neutral sector, in principle, allows for several neutral Higgs pair production channels at a muon collider. However, the underlying gauge structure and CP conservation severely restricts the allowed interactions. In particular, at tree level, neutral Higgs pair production proceeds exclusively through the associated $AH$ channel, while $HH$ and $AA$ final states are forbidden. A closer inspection of the Higgs-gauge couplings, summarized in Table~\ref{tab:couplings}, further constrains the phenomenology. In the alignment limit, all $AHZ$ couplings involving the SM-like Higgs boson vanish identically, rendering channels such as $\mu^+\mu^- \to Z^\ast \to A_i H_1$ inaccessible. The remaining $AHZ$ couplings exhibit a complementary structure, wherein the enhancement of one coupling is accompanied by the suppression of the other. Furthermore, an analysis of the allowed parameter space for the mixing angles $\gamma_1$ and $\alpha_3$, shown in Fig.~\ref{fig:mH2-al1-angles} and discussed in Sec.~\ref{sec:alignment}, indicates a preference for large mixing angles. As a consequence, the couplings $A_2H_2Z$ and $A_3H_3Z$ are expected to dominate over the corresponding cross-channel couplings, making these channels particularly promising for collider studies.
\begin{figure}[h!]
    \centering
    \includegraphics[scale=1.0]{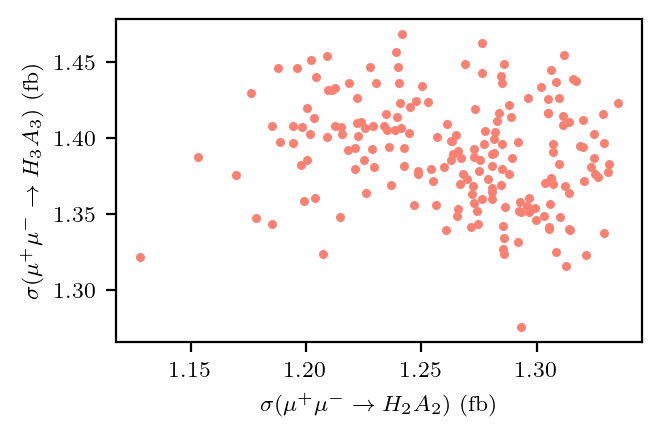}    
    \caption{Scatter plot showing the pair production cross section of $H_2A_2$ on the x-axis, and $H_3A_3$ on the y-axis for benchmark points consistent with all theoretical and experimental bounds.}
    \label{fig:xs}
\end{figure}

\begin{figure}[h!]
    \centering
    \includegraphics[scale=0.8]{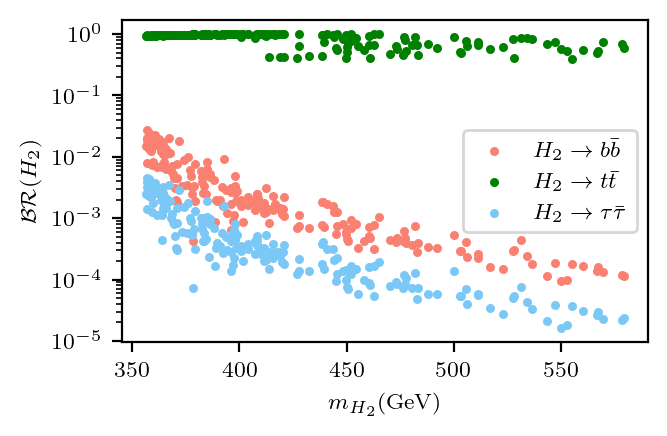} 
    \includegraphics[scale=0.8]{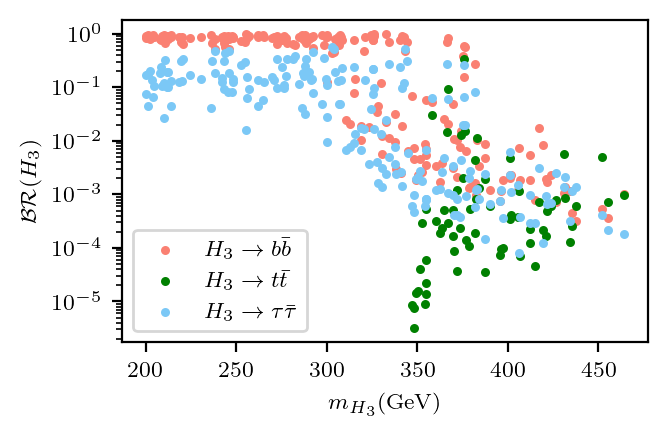}
    \includegraphics[scale=0.8]{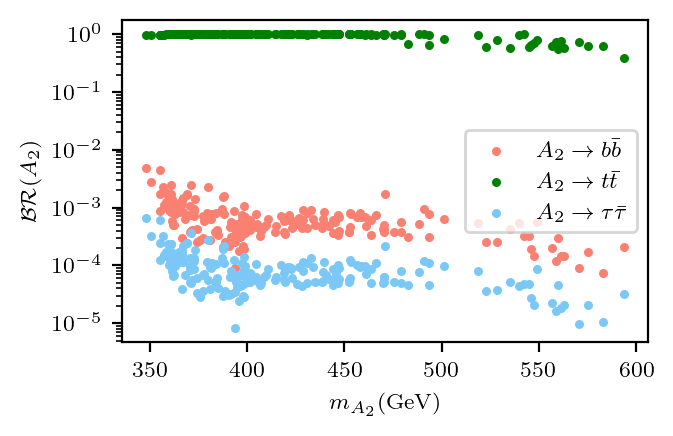}
    \includegraphics[scale=0.8]{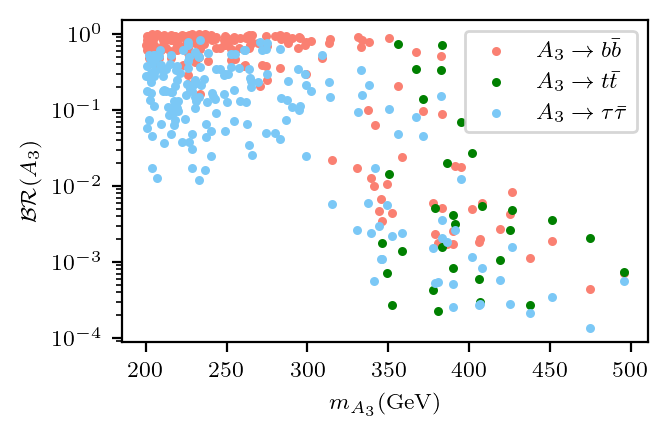}
    \caption{Branching Ratio plots of heavy Higgs}
    \label{fig:br}
\end{figure}

In Fig.~\ref{fig:xs}, we present the production cross sections for the $H_2A_2$ and $H_3A_3$ pairs via $\mu^+\mu^-$ annihilation for all benchmark scenarios that satisfy both theoretical and experimental constraints. As evident from the figure, the cross sections cluster within a narrow range, indicating that the different benchmark points are not expected to exhibit significant phenomenological differences at the production level. Consequently, the selection of representative benchmark points for a detailed collider analysis is primarily guided by the decay properties of the heavy Higgs states. In particular, we focus on benchmark scenarios that yield sizable branching fractions and thus maximize the total signal rate after decays. Motivated by this consideration, we examine the decay patterns of all heavy Higgs bosons in Fig.~\ref{fig:br}.

The Yukawa couplings of the heavy Higgs bosons are dominantly enhanced for third-generation fermions. Accordingly, in Fig.~\ref{fig:br}, we present the branching ratios of the heavy neutral Higgs states into $b\bar{b}$, $t\bar{t}$, and $\tau^+\tau^-$. The figure shows that, being heavier, $H_2$ and $A_2$ predominantly decay into top-quark pairs over most of the allowed parameter space. However, final states involving $t\bar{t}$ suffer from significant combinatorial ambiguities, rendering an accurate reconstruction of the parent Higgs invariant mass challenging. In view of this, we focus our collider analysis on the $H_3A_3$ production channel and consider two representative decay topologies: the fully hadronic $b\bar{b}b\bar{b}$ final state and the mixed $b\bar{b}t\bar{t}$ final state. These channels strike a balance between sizable branching fractions and experimentally tractable event reconstruction, making them particularly suitable for a detailed phenomenological study at a muon collider.
\begin{table}[h!]
    \centering
    \resizebox{\textwidth}{!}{
    \begin{tabular}{l l l l l l l l l l l l l l l}
    \toprule[1pt]
        BP & \ \ \ $mH_1$ & \ \ \  $mH_2$ & \ \ \ $mH_3$ & \ \ \ $mA_2$ & \ \ \ $mA_3$ & \ \ \ $mH^\pm_2$ & \ \ \ $mH^\pm_3$ & \ \ \ $\tan \beta_1$ & \ \ \ $\tan \beta_2$ & \ \ \ $\tan \alpha_1$ & \ \ \ $\tan \alpha_2$ & \ \ \ $\tan \alpha_3$ & \ \ \ $\tan \gamma_1$ & \ \ \ $\tan \gamma_2$ \\
        \midrule[1pt]
        BP1 & \ \ \ 125 & \ \ \ 506.10 & \ \ \ 230.48 & \ \ \ 396.90 & \ \ \ 232.85 & \ \ \ 395.58 & \ \ \ 318.13 & \ \ \ 1.04 & \ \ \ 0.90 & \ \ \ 1.04 & \ \ \ 0.90 & \ \ \ 94.40 & \ \ \ 98.85 & \ \ \ 66.54 \\ 
        BP2 & \ \ \ 125 & \ \ \ 444.32 & \ \ \ 204.69 & \ \ \ 366.75 & \ \ \ 299.00 & \ \ \ 405.94 & \ \ \ 353.65 & \ \ \ 0.85 & \ \ \ 1.12 & \ \ \ 0.85 & \ \ \ 1.12 & \ \ \ 82.98 & \ \ \ 36.02 & \ \ \ 58.64 \\
        BP3 & \ \ \ 125 & \ \ \ 411.88 & \ \ \ 265.43  & \ \ \ 434.92 & \ \ \ 355.79 & \ \ \ 413.30 & \ \ \ 405.45 & \ \ \ 1.05 & \ \ \ 1.03 & \ \ \ 1.05 & \ \ \ 1.03 & \ \ \ 16.12 & \ \ \ 12.75 & \ \ \ 28.98 \\ 
    \bottomrule[1pt]
    \bottomrule[1pt]
    \end{tabular}}
    \caption{Benchmark points used in the current work to study the process $\mu^+\mu^-\to A_i H_j$, where $H_j$ is a non-SM heavy Higgs boson. }
    \label{tab:benchmarkpoint}
\end{table}

We focus our attention on three representative benchmark scenarios, summarized in Table~\ref{tab:benchmarkpoint}. Benchmark Point 1 (BP1) corresponds to a fully hadronic final state with both Higgs bosons decaying into $b$-quark pairs, and features a nearly degenerate CP-even and CP-odd Higgs mass spectrum. Benchmark Point 2 (BP2) also leads to a fully hadronic $b\bar{b}b\bar{b}$ final state, but with the two scalar masses separated by approximately $100~\mathrm{GeV}$. Benchmark Point 3 (BP3) represents a mixed decay topology, where $H_3$ decays into a $b\bar{b}$ pair while $A_3$ decays into a $t\bar{t}$ pair. The corresponding total signal cross sections for these benchmark points are displayed in Table~\ref{tab:signal}.
\begin{table}[H]
    \centering
    \resizebox{\textwidth}{!}{
    \begin{tabular}{l l l l l l}
    \toprule[1pt]
        Channel & \ \ \ \ \ \ \ Final State & \ \ \ \ \ \ \ Cross Section (in fb) & \ \ \ \ \ \ \ $\mathcal{BR}$ & \ \ \ \ \ \ \ $\mathcal{BR}$ & \ \ \ \ \ \ \ Total xs (in fb) \\
        \midrule[1pt]
        $\mu^+\mu^- \rightarrow H_3 A_3 $ (BP1) & \ \ \ \ \ \ \ $b \Bar{b} b \Bar{b}$ & \ \ \ \ \ \ \ $1.459$ & \ \ \ \ \ \ \ $\mathcal{BR}(H_3 \to b \bar{b})=0.8568$ & \ \ \ \ \ \ \ $\mathcal{BR}(A_3 \to b \bar{b})=0.8496$ & \ \ \ \ \ \ \ $1.06$ \\
        $ \mu^+\mu^- \rightarrow H_3 A_3 $ (BP2) & \ \ \ \ \ \ \ $b \Bar{b} b \Bar{b}$ & \ \ \ \ \ \ \ $1.435$ & \ \ \ \ \ \ \ $\mathcal{BR}(H_3 \to b \bar{b})=0.933$ & \ \ \ \ \ \ \ $\mathcal{BR}(A_3 \to b \bar{b})=0.298$ & \ \ \ \ \ \ \ $0.399$ \\
        $ \mu^+\mu^- \rightarrow H_3 A_3 $ (BP3) & \ \ \ \ \ \ \ $b \Bar{b} t \Bar{t}$ & \ \ \ \ \ \ \ $1.355$ & \ \ \ \ \ \ \ $\mathcal{BR}(H_3 \to b \bar{b})=0.8682$ & \ \ \ \ \ \ \ $\mathcal{BR}(A_3 \to t \bar{t})=0.7335$ & \ \ \ \ \ \ \ $0.382$ \\
    \bottomrule[1pt]
    \bottomrule[1pt]
    \end{tabular}}
    \caption{Signal cross section}
    \label{tab:signal}
\end{table}

Event generation for this analysis was performed using the \texttt{MADGRAPH5\_aMC@NLO} framework~\cite{Alwall:2011uj}. The Standard Model background processes considered are summarized in Table~\ref{tab:background} and were simulated at a center-of-mass energy of $3~\mathrm{TeV}$ using the default SM implementation available within the \texttt{MADGRAPH} repository. Signal events were generated using a model file constructed with the \texttt{SARAH} package~\cite{Staub:2008uz}, while the corresponding mass spectrum and model parameters were provided by \texttt{SPheno}~\cite{Porod:2003um}. The parton-level events produced by \texttt{MADGRAPH} were subsequently interfaced with \texttt{PYTHIA}~\cite{Bierlich:2022pfr} for parton showering and hadronization. Detector effects were modeled using \texttt{DELPHES}~\cite{deFavereau:2013fsa}, and final-state object reconstruction was carried out with \texttt{MADANALYSIS5}~\cite{Conte:2012fm}, which was also employed to perform the cut-and-count analysis discussed in the following section. The production cross sections were computed using \texttt{Mg5\_aMC\_v\_3.6.6}, with Initial State Radiation (ISR) included. 
\begin{table}[h!]
    \centering
    \begin{tabular}{l l l}
    \toprule[1pt]
        Channel & \ \ \ \ \ \ \ Final State & \ \ \ \ \ \ \ Cross Section (in fb) \\
        \midrule[1pt]
        $\mu^+\mu^- \rightarrow b \Bar{b} b \Bar{b} $ & \ \ \ \ \ \ \ $b \Bar{b} b \Bar{b}$ & \ \ \ \ \ \ \ $1.297$  \\
        $\mu^+\mu^- \rightarrow b \Bar{b} b \Bar{b} b \Bar{b}$ & \ \ \ \ \ \ \ $b \Bar{b} b \Bar{b} b \Bar{b}$ & \ \ \ \ \ \ \ $0.004915$  \\
        $\mu^+\mu^- \rightarrow b \Bar{b} t \Bar{t} $ & \ \ \ \ \ \ \ $b \Bar{b} b \Bar{b} + 4j$ & \ \ \ \ \ \ \ $0.3924$  \\
        $\mu^+\mu^- \rightarrow b \Bar{b} t \Bar{t} $ & \ \ \ \ \ \ \ $b \Bar{b} b \Bar{b} + l^+l^-$ & \ \ \ \ \ \ \ $0.043$  \\
        $\mu^+\mu^- \rightarrow t \Bar{t} t \Bar{t} $ & \ \ \ \ \ \ \ $b \Bar{b} b \Bar{b} + 8j$ & \ \ \ \ \ \ \ $0.003157$   \\
        $\mu^+\mu^- \rightarrow t \Bar{t} t \Bar{t} $ & \ \ \ \ \ \ \ $b \Bar{b} b \Bar{b} + 6j + l^+$ & \ \ \ \ \ \ \ $0.005604$   \\
        $\mu^+\mu^- \rightarrow t \Bar{t} t \Bar{t} $ & \ \ \ \ \ \ \ $b \Bar{b} b \Bar{b} + 6j + l^-$ & \ \ \ \ \ \ \ $0.005614$   \\
        $\mu^+\mu^- \rightarrow t \Bar{t} t \Bar{t} $ & \ \ \ \ \ \ \ $b \Bar{b} b \Bar{b} + 4j + l^+l^-$ & \ \ \ \ \ \ \ $0.009932$   \\
        $\mu^+\mu^- \rightarrow t \Bar{t} t \Bar{t} $ & \ \ \ \ \ \ \ $b \Bar{b} b \Bar{b} + 2j + l^+l^-l^+$ & \ \ \ \ \ \ \ $0.000622$   \\
        $\mu^+\mu^- \rightarrow t \Bar{t} t \Bar{t} $ & \ \ \ \ \ \ \ $b \Bar{b} b \Bar{b} + 2j + l^+l^-l^-$ & \ \ \ \ \ \ \ $0.000623$   \\
        $\mu^+\mu^- \rightarrow t \Bar{t} t \Bar{t} $ & \ \ \ \ \ \ \ $b \Bar{b} b \Bar{b} + l^+l^-l^+l^-$ & \ \ \ \ \ \ \ $0.000039$   \\
        $\mu^+\mu^- \rightarrow jjjj $ & \ \ \ \ \ \ \ $4j$ & \ \ \ \ \ \ \  $150.4$ \\
    \bottomrule[1pt]
    \bottomrule[1pt]
    \end{tabular}
    \caption{The dominant SM backgrounds that are relevant in the collider study of $A_iH_j$ production with subsequent decays to bottom and/or top quark pairs.}
    \label{tab:background}
\end{table}
\subsection{Analysis and Results}
\begin{figure}[h!]
        \includegraphics[scale=0.7]{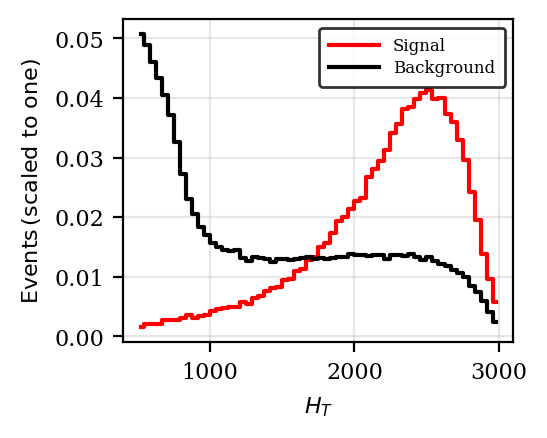}
        \includegraphics[scale=0.7]{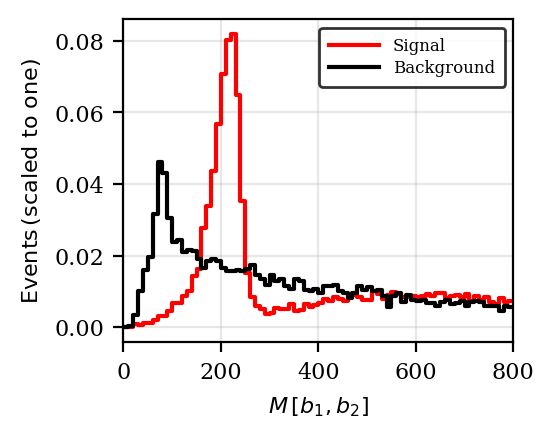}
        \includegraphics[scale=0.7]{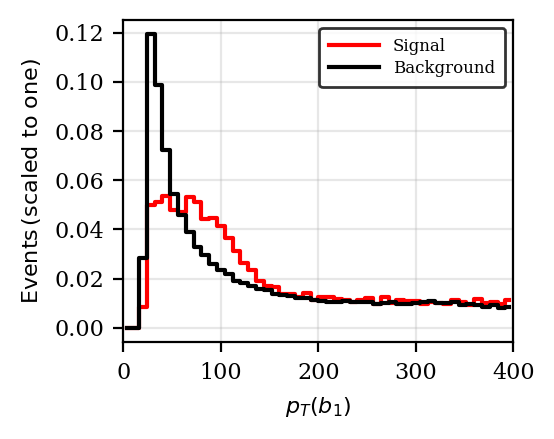}
        \caption{The various kinematic observables for BP1}
         \label{fig:BP1}
\end{figure}
The cut flow for Benchmark Point 1 (BP1) is presented in Table~\ref{tab:BP1}, with all results quoted for an integrated luminosity of $1000 \ \textrm{fb}^{-1}$. This benchmark corresponds to a nearly degenerate mass spectrum, with the CP-even scalar and the CP-odd scalar having masses of $230$ GeV and $232$ GeV respectively. Both scalars decay dominantly into bottom-quark pairs, leading to a fully hadronic $4b$ final state. We begin the analysis by imposing a basic signal identification requirement of at least two b-tagged jets, $N(b) \geq 2$. This cut significantly suppresses the SM backgrounds while retaining a sizable fraction of the signal. Although the signal contains four $b$-jets at the parton level, detector effects and jet merging in the boosted regime can reduce the reconstructed $b$-jet multiplicity, motivating the relatively loose requirement at this stage. Given the large center-of-mass energy of the muon collider, the produced scalar pair is highly boosted and as a result, the decay products inherit large transverse momenta. This feature is exploited by imposing a lower bound on the transverse momentum of the leading b-jet, $p_T(b_1) > 200 \ \textrm{GeV}$. This cut efficiently removes soft QCD backgrounds while preserving most of the signal events, leading to a modest improvement in the signal significance. To further characterize the energetic nature of the signal, we impose a cut on the scalar sum of transverse momenta of all reconstructed jets, $H_T > 1000 \ \textrm{GeV}$. Owing to the presence of two heavy resonances in the final state, signal events populate the high-$H_T$ region, whereas the background falls off rapidly. This cut provides additional background suppression with minimal signal loss, resulting in a slight enhancement of the statistical sensitivity. Finally, we reconstruct the invariant mass of the two leading b-jets and require it to lie within the window $160<M(b_1,b_2)<260$. This mass window is chosen to capture the decay products of the nearly degenerate $H_3$ and $A_3$ states. After applying all selection cuts, the analysis achieves a final statistical significance close to the $5 \sigma$ discovery threshold for BP1. The distributions of the key kinematic observables used in the analysis are presented in Figure \ref{fig:BP1}.
\begin{table}[H]
    \centering
    \begin{tabular}{l l l l}
    \toprule[1pt]
        Cut & \ \ \ \ \ \ \ Signal & \ \ \ \ \ \ \ Background & \ \ \ \ \ \ \ S vs B \\
        \midrule[1pt]
        Initial & \ \ \ \ \ \ \ $1060$ & \ \ \ \ \ \ \ $152162$ & \ \ \ \ \ \ \ $-$ \\
        $N(b)>=2$ & \ \ \ \ \ \ \ $316$ & \ \ \ \ \ \ \ $2448$ & \ \ \ \ \ \ \ $6.02$ \\
        $p_T(b_1
        )>200$ & \ \ \ \ \ \ \ $292$ & \ \ \ \ \ \ \ $1784$ & \ \ \ \ \ \ \ $6.41$ \\
        $H_T>1000$ & \ \ \ \ \ \ \ $268$ & \ \ \ \ \ \ \ $1392$ & \ \ \ \ \ \ \ $6.58$ \\
        $160<M(b_1,b_2)<260$ & \ \ \ \ \ \ \ $80$ & \ \ \ \ \ \ \ $169$ & \ \ \ \ \ \ \ $5.07$ \\
    \bottomrule[1pt]
    \bottomrule[1pt]
    \end{tabular}
    \caption{Cutflow chart for the analysis of BP1.}
    \label{tab:BP1}
\end{table}
\begin{figure}[h!]
        \includegraphics[scale=0.7]{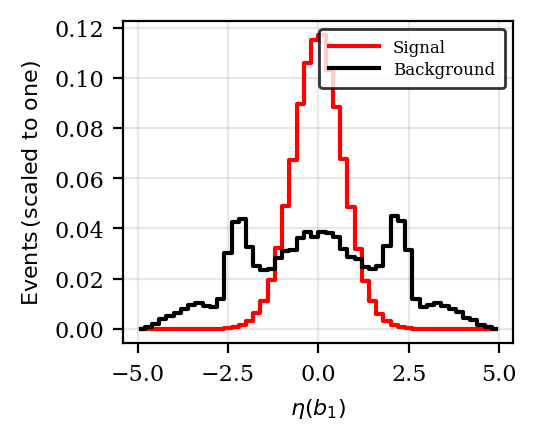}
        \includegraphics[scale=0.7]{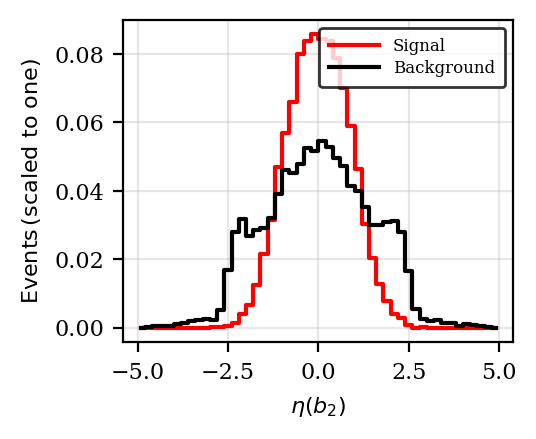}
        \includegraphics[scale=0.7]{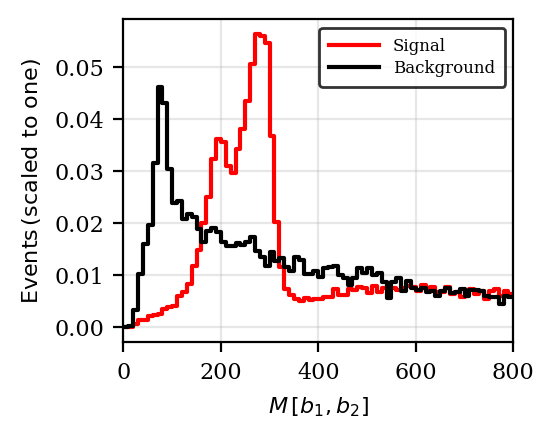}
        \caption{The various kinematic observables for BP2}
         \label{fig:BP2-initial}
\end{figure}
We now turn to Benchmark Point 2 (BP2), for which the cut flow is summarized in Table~\ref{tab:BP2}. The analysis is performed at an integrated luminosity of $4000 \ \textrm{fb}^{-1}$. In this scenario, the CP-even scalar $H_3$ and the CP-odd scalar $A_3$ are non-degenerate, with masses of $204$ GeV and $299$ GeV, respectively, and both decay hadronically into bottom-quark pairs, resulting in a $4b$ final state. The kinematic distributions prior to the application of selection cuts are shown in Fig.~\ref{fig:BP2-initial}. In particular, the invariant mass distribution of the two leading $b$-tagged jets exhibits two distinct resonant structures around $200$ GeV and $300$ GeV, corresponding to the decays of $H_3$ and $A_3$, respectively. This mass separation motivates a cut-based strategy that exploits both kinematic and angular correlations to identify the correct jet pairing. We begin the event selection by requiring at least two $b$-tagged jets, $N(b) \geq 2$, followed by a lower bound on the transverse momentum of the leading $b$-jet $p_T(b_1) > 250$ GeV. These requirements efficiently suppress the soft SM background while retaining a large fraction of the signal, reflecting the highly boosted nature of the scalar pair produced at a multi-TeV muon collider. To further reduce the background contribution from forward events, we impose centrality cuts on the two leading $b$-jets, requiring $|\eta(b_1)|<1.3$ and $|\eta(b_2)|<1.5$. 
\begin{figure}[h!]             
        \includegraphics[scale=0.65]{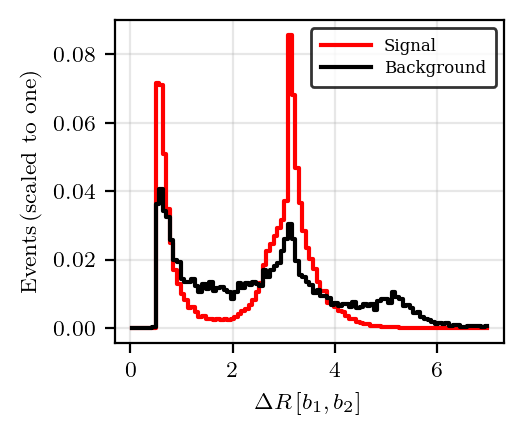} 
        \includegraphics[scale=0.65]{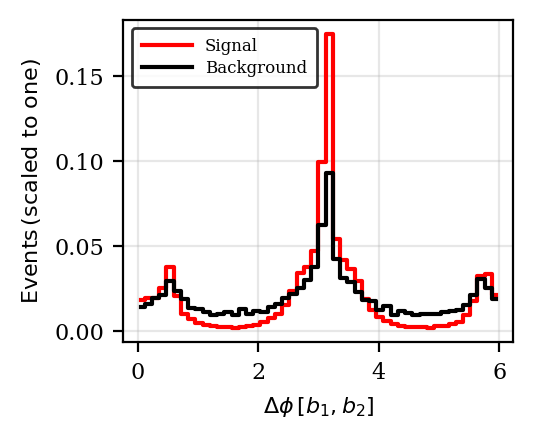}
        \includegraphics[scale=0.65]{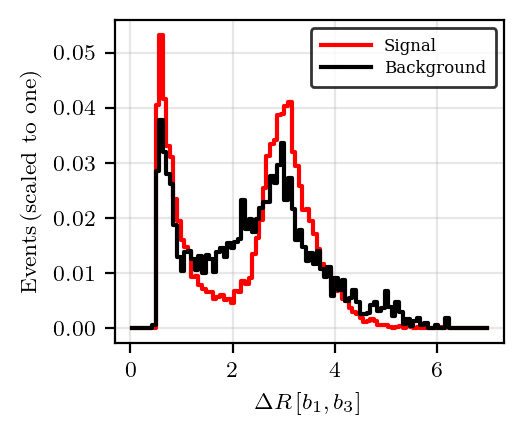}  
        \includegraphics[scale=0.65]{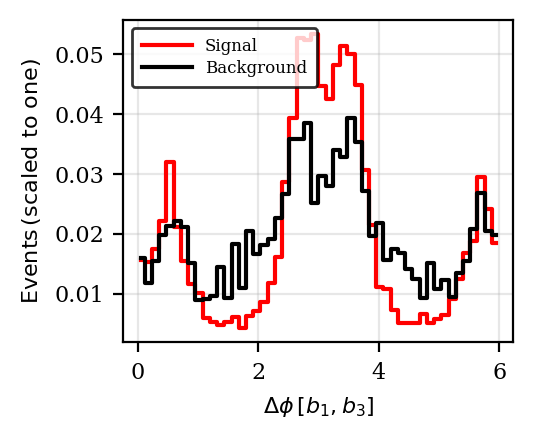} 
        \caption{The various kinematic variables for BP2 before applying the $\Delta R$ cut} 
        \label{fig:BP2-angles} 
\end{figure}

\begin{figure}[h!]             
        \includegraphics[scale=0.64]{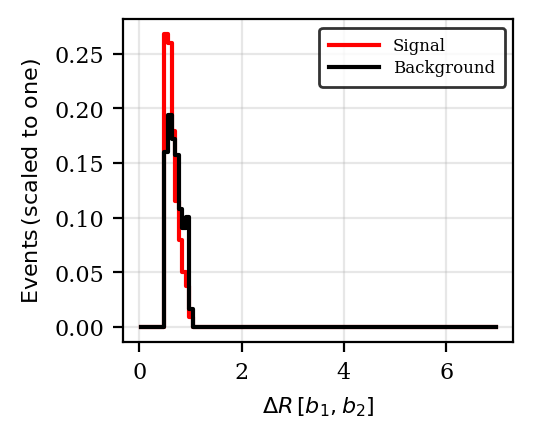} 
        \includegraphics[scale=0.64]{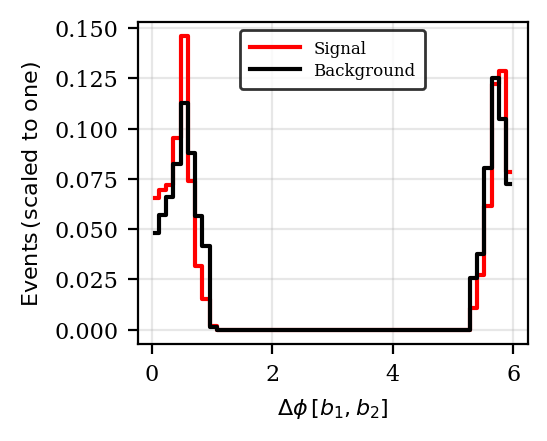} 
        \includegraphics[scale=0.64]{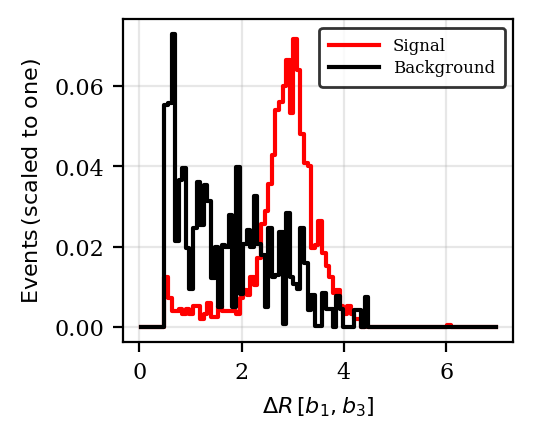} 
        \includegraphics[scale=0.64]{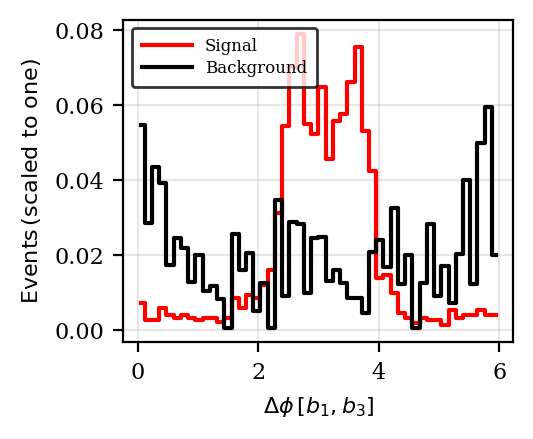} 
        \caption{The various kinematic variables for BP2 after applying the $\Delta R$ cut} 
        \label{fig:BP2-angles-cut} 
\end{figure}

We next exploit angular correlations among the $b$-jets to identify pairs originating from the same Higgs boson. The distributions of the angular variables $\Delta R$ and $\Delta \phi$ for the jet pairs $(b_1,b_2)$ and $(b_1,b_3)$ prior to imposing angular cut, are shown in Fig.~\ref{fig:BP2-angles}. The $\Delta R$ distributions exhibit a characteristic double-peak structure, while the corresponding azimuthal angle distributions show peaks near $\Delta \phi \simeq 0, 2\pi \ \textrm{and} \ \pi$, reflecting the presence of both collimated and back-to-back jet configurations. Upon imposing the angular selection $0.4 < \Delta R(b_1,b_2)<1.0$ , events corresponding to the collimated jet configuration are preferentially selected. This selection is corroborated by the corresponding $\Delta \phi (b_1,b_2)$ distribution, which peaks near $0$ and $2 \pi$, indicating that the two jets are produced with a small angular separation in the transverse plane. In contrast, the remaining events in the $(b_1,b_3)$ pairing populate the larger $\Delta R$ region and peak near $\Delta \phi \simeq \pi$, consistent with jets originating from different parent particles. The distributions of these angular variables after imposing the angular cut, are shown in Figure \ref{fig:BP2-angles-cut}. This behavior strongly suggests that the $(b_1,b_2)$ jet pair selected by the angular cut has a higher probability of originating from the decay of a single Higgs boson. Consequently, we apply the angular selection prior to reconstructing the invariant mass. Finally, we impose the mass window $210 < M(b_1,b_2) < 320$ after which the analysis achieves a final significance exceeding the $5 \sigma$, the discovery threshold for BP2.
\begin{table}[H]
    \centering
    \begin{tabular}{l l l l}
    \toprule[1pt]
        Cut & \ \ \ \ \ \ \ Signal & \ \ \ \ \ \ \ Background & \ \ \ \ \ \ \ S vs B \\
        \midrule[1pt]
        Initial & \ \ \ \ \ \ \ $1596$ & \ \ \ \ \ \ \ $608651$ & \ \ \ \ \ \ \ $-$ \\
        $N(b)>=2$ & \ \ \ \ \ \ \ $554$ & \ \ \ \ \ \ \ $9792$ & \ \ \ \ \ \ \ $5.45$ \\
        $p_T(b_1
        )>250$ & \ \ \ \ \ \ \ $503$ & \ \ \ \ \ \ \ $6492$ & \ \ \ \ \ \ \ $6.01$ \\
        $-1.3<\eta(b_1)<1.3$ & \ \ \ \ \ \ \ $466$ & \ \ \ \ \ \ \ $4728$ & \ \ \ \ \ \ \ $6.47$ \\
        $-1.5<\eta(b_2)<1.5$ & \ \ \ \ \ \ \ $447$ & \ \ \ \ \ \ \ $4136$ & \ \ \ \ \ \ \ $6.60$ \\
        $0.4<\Delta R(b_1,b_2)<1.0$ & \ \ \ \ \ \ \ $142$ & \ \ \ \ \ \ \ $940$ & \ \ \ \ \ \ \ $4.33$ \\
        $210<M(b_1,b_2)<320$ & \ \ \ \ \ \ \ $104$ & \ \ \ \ \ \ \ $304$ & \ \ \ \ \ \ \ $5.16$ \\
    \bottomrule[1pt]
    \bottomrule[1pt]
    \end{tabular}
    \caption{Cutflow chart for the analysis of BP2.}
    \label{tab:BP2}
\end{table}
\begin{figure}[h!]
       \includegraphics[scale=0.7]{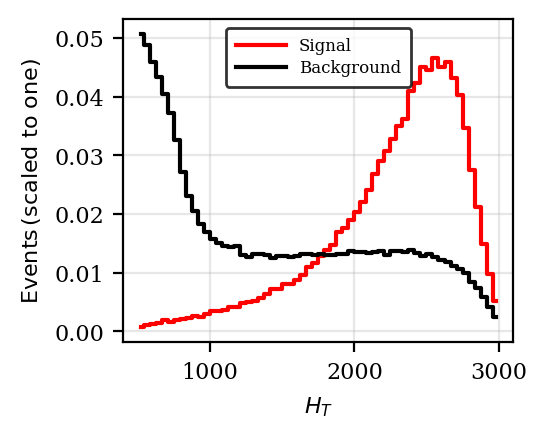}
        \includegraphics[scale=0.7]{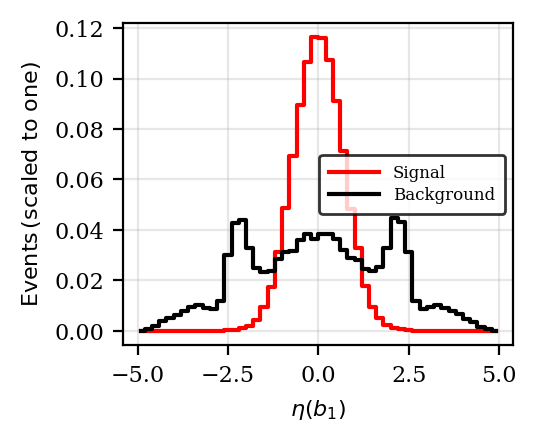}
        \includegraphics[scale=0.7]{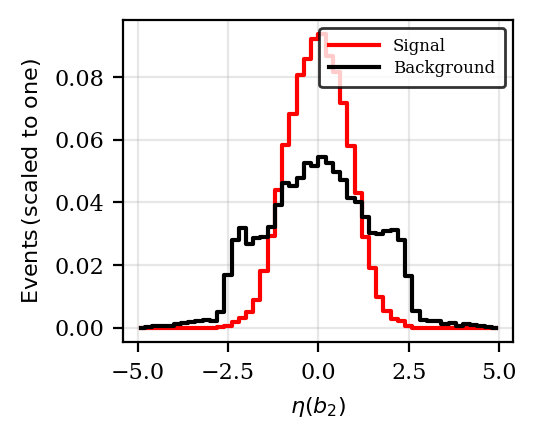}
        \includegraphics[scale=0.7]{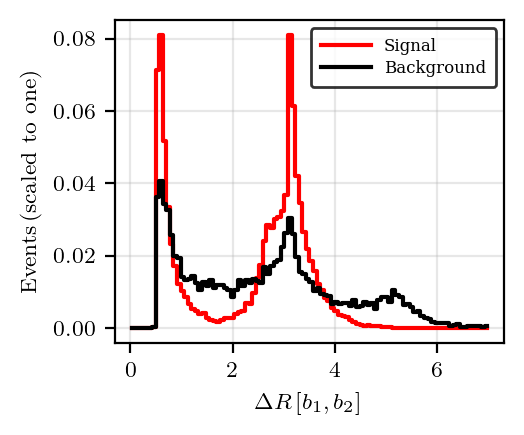}
        \includegraphics[scale=0.7]{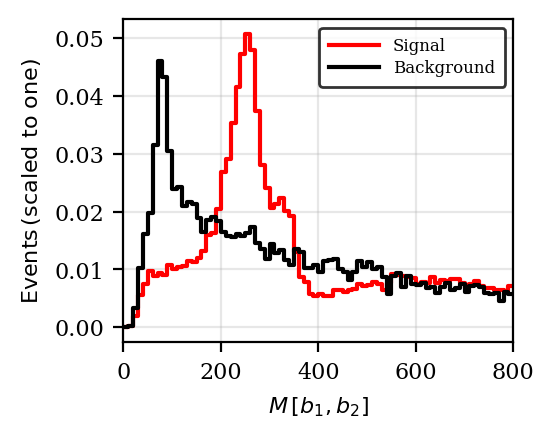}
        \caption{The relevant kinematic distributions used in the analysis of BP-3.}
        \label{fig:BP3}
\end{figure}
Finally, we turn to Benchmark Point 3 (BP3), whose cut flow is summarized in Table~\ref{tab:BP3}. In this scenario, the CP-even scalar $H_3$, with mass $265$ GeV decays into a bottom-quark pair, while the CP-odd scalar $A_3$ decays into a top-quark pair. The subsequent decay of the top quarks, $t \to Wb$, with the $W$ bosons decaying hadronically, leads to a mixed hadronic final state characterized by multiple $b$-jets and light-flavor jets. The relevant kinematic distributions for BP3 are shown in Fig.~\ref{fig:BP3}. In particular, the invariant mass distribution of the two leading $b$-tagged jets exhibits a resonant structure around $260$ GeV, corresponding to the decay of the CP-even Higgs boson. Owing to the complexity of the final state and the presence of additional jets from top-quark decays, we restrict our reconstruction strategy to the $h \to b \bar{b}$ channel. 
We begin the event selection by requiring at least two $b$-tagged jets, $N(b) \geq 2$, along with a minimum jet multiplicity, $N(j) \geq 2$, to ensure sensitivity to the hadronic activity arising from top-quark decays. To suppress backgrounds dominated by forward and soft jet production, we impose centrality cuts on the two leading $b$-jets requiring  $|\eta(b_1)|<1.1$ and $|\eta(b_2)|<1.5$, followed by a requirement on the scalar sum of transverse momenta $H_T > 1100$. These cuts efficiently reduce the background while retaining a substantial fraction of the signal. As in the BP2 analysis, we exploit angular correlations among the $b$-jets to enhance the probability of selecting the correct Higgs decay products. In particular, we impose the angular requirement $0.4 < \Delta R(b_1,b_2)<1.0$, which preferentially selects collimated $b$-jet pairs consistent with the decay of a boosted scalar, while suppressing combinatorial pairings involving $b$-jets from top-quark decays. Finally, we reconstruct the invariant mass of the selected $b$-jet pair and impose the mass window $190 < M(b_1,b_2) < 320$ centered around the CP-even Higgs mass. After applying all selection criteria, the analysis achieves a final statistical significance exceeding the $5 \sigma$ discovery threshold for the BP3 for an integrated luminosity of $4000 \ \textrm{fb}^{-1}$.
\begin{table}[H]
    \centering
    \begin{tabular}{l l l l}
    \toprule[1pt]
        Cut & \ \ \ \ \ \ \ Signal & \ \ \ \ \ \ \ Background & \ \ \ \ \ \ \ S vs B \\
        \midrule[1pt]
        Initial & \ \ \ \ \ \ \ $1528$ & \ \ \ \ \ \ \ $608651$ & \ \ \ \ \ \ \ $-$ \\
        $N(b)>=2$ & \ \ \ \ \ \ \ $603$ & \ \ \ \ \ \ \ $9792$ & \ \ \ \ \ \ \ $5.92$ \\
        $N(j)>=2$ & \ \ \ \ \ \ \ $603$ & \ \ \ \ \ \ \ $9792$ & \ \ \ \ \ \ \ $5.92$ \\
        $-1.1<\eta(b_1)<1.1$ & \ \ \ \ \ \ \ $525$ & \ \ \ \ \ \ \ $5391$ & \ \ \ \ \ \ \ $6.82$ \\
        $-1.5<\eta(b_2)<1.5$ & \ \ \ \ \ \ \ $509$ & \ \ \ \ \ \ \ $4669$ & \ \ \ \ \ \ \ $7.07$ \\
        $H_T>1100$ & \ \ \ \ \ \ \ $498$ & \ \ \ \ \ \ \ $4378$ & \ \ \ \ \ \ \ $7.13$ \\
        $0.4<\Delta R(b_1b_2)<1.0$ & \ \ \ \ \ \ \ $156$ & \ \ \ \ \ \ \ $1100$ & \ \ \ \ \ \ \ $4.40$ \\
        $190<M(b_1b_2)<320$ & \ \ \ \ \ \ \ $114$ & \ \ \ \ \ \ \ $345$ & \ \ \ \ \ \ \ $5.34$ \\
    \bottomrule[1pt]
    \bottomrule[1pt]
    \end{tabular}
    \caption{Cutflow chart for the analysis of BP3.}
    \label{tab:BP3}
\end{table}
\section{Discussion and Conclusions}
\label{sec:conclusions}
In this work, we have investigated the collider phenomenology of the $Z_3$-symmetric Three Higgs Doublet Model with a Type-Z Yukawa structure at a future multi-TeV muon collider. The model features a rich scalar spectrum consisting of three CP-even Higgs bosons, two CP-odd Higgs bosons, and a pair of charged Higgs states, leading to a variety of possible production and decay topologies that are qualitatively distinct from those of the Two Higgs Doublet Model.

Building on earlier analysis of the $Z_3$-symmetric 3HDM parameter space, in which theoretical consistency requirements such as vacuum stability, perturbative unitarity, and perturbativity were combined with experimental constraints from Higgs signal strengths, direct searches, electroweak precision observables, and flavor data, we adopt the viable regions obtained in the alignment limit where the lightest CP-even scalar reproduces the observed 125 GeV Higgs boson. Within this framework, neutral scalar pair production at a muon collider is shown to proceed predominantly through the associated $AH$ channel mediated by an off-shell $Z$ boson, while $HH$ and $AA$ production is forbidden at tree level. Furthermore, the structure of scalar mixing angles induces a complementary pattern in the $AHZ$ couplings, resulting in the dominance of the $A_2H_2$ and $A_3H_3$ production channels.

Motivated by these features, we performed a detailed detector-level analysis of the $H_3A_3$ production channel at a center-of-mass energy of $\sqrt{s}=3$ TeV. Benchmark points consistent with all constraints were selected based on their decay patterns, yielding two experimentally promising final states: the fully hadronic $b\bar{b}b\bar{b}$ topology and the mixed $b\bar{b}t\bar{t}$ topology.  A cut-and-count strategy exploiting boosted kinematics, jet multiplicities, angular correlations, and invariant mass reconstruction was employed to enhance signal sensitivity. Our analysis shows that the muon collider provides excellent discovery prospects for heavy neutral scalars predicted by the 3HDM. For representative benchmark scenarios, discovery-level statistical significance can be achieved with integrated luminosities of $\mathcal{O}(1\text{--}4~\mathrm{ab}^{-1})$, depending on the mass hierarchy and decay topology. The clean experimental environment and the availability of the full center-of-mass energy for the hard scattering process play a crucial role in enabling efficient reconstruction of heavy scalar resonances and in suppressing QCD backgrounds that typically limit hadron-collider searches.

The results presented here highlight several distinctive features of the 3HDM that could be probed at a muon collider. In particular, the presence of multiple CP-even and CP-odd states leads to characteristic associated-production patterns and correlated decay signatures that can serve as diagnostic indicators of multi-doublet Higgs sectors beyond the 2HDM. Moreover, the democratic Yukawa structure allows for enhanced couplings to third-generation fermions, resulting in sizable branching fractions into $b\bar{b}$ and $t\bar{t}$ final states that are experimentally accessible.

In conclusion, our study demonstrates that a future muon collider offers a powerful and complementary probe of extended Higgs sectors, with the capability to discover heavy neutral scalars predicted by the $Z_3$-symmetric 3HDM over a significant region of parameter space. These results reinforce the importance of muon collider programs in the search for physics beyond the Standard Model and highlight the rich phenomenology associated with multi-Higgs doublet frameworks.

\begin{acknowledgments}
A.K. acknowledges the support from the Director's Fellowship at IIT Gandhinagar.
\end{acknowledgments}

\bibliography{bibliography}

\end{document}